\documentstyle[preprint,aps]{revtex}
\begin{document}
\draft
\title{{\bf Singlet Magnetism in Heavy Fermions}}
\author{\large Victor Barzykin}
\address{Department of Physics, University of Illinois at Urbana-Champaign, \\
1110 West Green Street, Urbana, Illinois 61801-3080, USA}
\author{\large Lev P. Gor'kov}
\address{National High Magnetic Field Laboratory,
Florida State University, \\
1800 East Paul Dirac Dr., Tallahassee, FL 32306-4005, USA\\
and \\
L.D.Landau Institute for Theoretical Physics, Moscow, Russia}
\maketitle
\begin{abstract}
We consider singlet magnetism for the uranium ions
in UPt$_3$ and URu$_2$Si$_2$
assuming that time-reversal symmetry is broken for
the {\em coherent state of intermediate valence}.
The relative weight of the two involved configurations should be different
for UPt$_3$ and URu$_2$Si$_2$.
If in UPt$_3$ the configuration $5f^1$
on the U-ion
prevails in the coherent state below the magnetic transition, the
magnetic moment would vanish for
the particular choice of the {\em ionic}
wave function. In case of
URu$_2$Si$_2$, the phase transition is
non-magnetic in the first approximation --
the magnetic moment arises from a small
admixture of a half-integer spin configuration.
\end{abstract}

\vspace{0.3in}
\narrowtext

Heavy Fermion (HF) compounds reveal enormously rich
magnetic properties in which both localized states
and conduction electrons are involved\cite{Mydosh}.
To be more specific, among the magnetic phases observed
in $U$-based materials
are  the antiferromagnetic (AF) N\'eel state in
U$_2$Zn$_{17}$ \cite{Ott}, a ``tiny-moment'' AF state in URu$_2$Si$_2$
\cite{Mason,Walker} and UPt$_3$
($\langle \mu \rangle \sim 10^{-2} \mu_B$) and both
quadrupolar and ``tiny-moment'' phases in UPd$_3$ \cite{Buyers,McEwen}.
This diversity may originate from the fine energy structure
of 10-100K for the uranium multiplets, somehow supplemented by
strong electron correlations. In what follows
we concentrate on the ``tiny moment'' antiferromagnetic transitions
in the HF compounds UPt$_3$ and URu$_2$Si$_2$. In view of the
smallness of the above energy scale we suggest that the mixed valency (MV)
phenomena may be involved.

While the small ordered antiferromagnetic moment is the
common feature for both compounds
the phase transitions in them are very different. Thus,
a huge specific heat jump is observed at the N\'eel transition
in URu$_2$Si$_2$($T_N \sim 17.5K$) and no specific heat anomaly was seen
in UPt$_3$. The well-defined excitations with large
$<0|J^z|1>$ ($ \sim 1.2 \mu_B$) \cite{Mason} matrix element
are detected in URu$_2$Si$_2$ by inelastic neutron
scattering in the ordered state and
again are absent in UPt$_3$\cite{UPt3}. In URu$_2$Si$_2$
the excitations below $T_N$ can be ascribed to the transitions
between two well-defined singlet
states of the Hund's rule $^3H_4$ multiplet for the $U^{4+}$ ion.
These excitations have a finite gap at zero temperature. However, as the
temperature is raised, the gap in the spin-wave spectrum reduces and
finally vanishes at the N\'eel point \cite{Mason:new}.
In our opinion, this signals the onset of large valency fluctuations
above the N\'eel temperature.
Although the standard singlet-singlet scheme  describes fairly well the
temperature behavior of the bulk
magnetic susceptibility and  the heat capacity \cite{Mason},
it fails to reconcile the small magnetic moment and
huge specific heat anomaly at the N\'eel temperature.
 A multipolar magnetic order of different symmetry as
the primary order parameter in URu$_2$Si$_2$
was also excluded from the symmetry analysis
of the neutron scattering measurements \cite{Walker}.
On the contrary, there is no specific heat jump
associated with the transition in UPt$_3$, and no well-defined
crystal field excitations were detected\cite{UPt3}.
The tiny moment $M^2 \propto |T - T_N|$ displays a mean field
behavior over a wide temperature range below $T_N \simeq 5 K$.

  We argue that the effects of mixed valence of the U-ions  may be
responsible for the properties of both compounds.
The coherent wave function
for the sub-system of uranium ions, which forms below the phase transition,
then involves the low-energy levels of both configurations.
We consider first UPt$_3$.
For this compound we assume that
$5f^2$ and $5f^1$ configurations are close in energy (on the scale of the
f-level width).
The degeneracy of either configuration is lifted by the
strong spin-orbit interaction and the onset of the crystal field
splitting, so that in the framework of the singlet magnetism scheme the
low-energy states of the U-ion are the ground state singlet of the $5f^2$
configuration and some excited state Kramers doublet $\Gamma_{ex}$ of the
$5f^1$ configuration. We choose the matrix element between the
two low-energy levels as the order parameter:
\begin{equation}
\label{orpar}
\Psi= \langle0| \Gamma_0 \rangle \langle \Gamma_{ex} \alpha_{ex}| 0 \rangle
\end{equation}
We analyse the symmetry properties of this order parameter in a
model-independent fashion by using the Landau theory of phase transitions.
Since the excited state of the uranium ions has a half-integer spin,
the symmetry analysis should involve the double group.  The double point group
has an additional element ($\tilde{E}$), a rotation by the angle $2 \pi$,
which doubles the number of the group elements. The wave function of
an integer spin is invariant with respect to this transformation, while
in case of half-integer spin it changes sign. Thus, since
$\Psi$ transforms according to the representation
$\tilde{\Gamma}_{ex} = \Gamma_0 \otimes \Gamma_{ex}$ under the action
of the point group transformations, it is odd with respect to
$\tilde{E}$.

To enumerate various magnetic symmetry options, we start directly from the
order parameter Eq.(\ref{orpar}) and deduce magnetic classes which may
develop below the N\'eel temperature.
This analysis can be done in a fashion rather similar to
the classification of superconducting classes
\cite{Volovik:Gorkov,Gorkov:review}
in the HF materials. The total point symmetry group in the
``paramagnetic phase'' is $G \otimes R \otimes U(1)$, where
$G$ is the crystal double point group, R is the time-reversal symmetry,
U(1) is the group of the gauge transformation. The
symmetry class is the subgroup of this group appearing
below the phase transition.
We refer the reader to Ref.\cite{Gorkov:review} for the detailed
description of this classification and the methods of
construction of symmetry classes.
In Ref. \cite{Volovik:Gorkov} the symmetry classes for the wave functions
of integer spin (i.e., even with respect to the double group
transformation $\tilde{E}$) were enumerated.
The matrix element Eq.(\ref{orpar}) requires consideration
of the symmetry classes for the spinorial wave functions, which,
as said above,
are odd with respect to $\tilde{E}$.
We have derived the new symmetry classes for the wave function
of half-integer spin. The classes relevant to the present case
(i.e. for the double groups $D'_4$ and $D'_6$) are listed in
Table \ref{table}.
The {\em magnetic class} (i.e., the point symmetry of spin correlators)
can be obtained from the {\em symmetry class} by
substituting unity for all the gauge factors.
Thus, for example, for the representation $\tilde{E}_3$ of
the group $D'_6$ the magnetic moment
is fully prohibited by the magnetic group for
the phases II and III.
As seen from Table \ref{table}, otherwise the magnetic moment should
always appear.
Here we indicate an interesting possibility
when the magnetic moment, although expected by
the symmetry of the magnetic group in Table \ref{table},
turns out to be equal to zero
for the particular ionic wave function.
A small magnetic moment can then be induced as a result of the interaction
with the genuine order parameter \cite{Gorkov}. This
case is of the most interest
to us in connection
with the above mentioned puzzles in UPt$_3$. We found one example of this
accidental cancellation, for the $5f^1$ configuration in the
group $D_6'$, which is considered below. No such cancellations were
found in case of $D_4'$ (URu$_2$Si$_2$).

\noindent
{\em $J=5/2$ multiplet of the $5f^1$ configuration in a $D'_6$ group}.

For $J=5/2$ the six-fold degeneracy is lifted
by the crystal field effects, leaving three
Kramers doublets: $\tilde{E}_1$ ($| \pm 1/2 \rangle$),
$\tilde{E}_2$ ($| \pm 5/2 \rangle$) and
$\tilde{E}_3$ ($| \pm 3/2 \rangle$).
The nontrivial magnetic states cannot be constructed from
the  wave functions of $\tilde{E}_1$.
For the $\tilde{E}_2$ representation
the wave function can be written in a form:
\begin{equation}
\Psi = \eta_1 \psi_1 + \eta_2 \psi_2,
\end{equation}
with $\psi_1 = | 5/2 \rangle$, $\psi_2 = | - 5/2 \rangle$. Making the
substitution:
\begin{eqnarray}
\eta_1 & = & u sin \theta e^{i \phi_1} \nonumber \\
\eta_2 & = & u cos \theta e^{i \phi_2},
\end{eqnarray}
we arrive to the following form of the Landau
functional:
\widetext
\begin{equation}
\label{Landau1}
F = - \alpha \tau u^2 + \beta_1 u^4 + \beta_2 u^4 sin^2 (2 \theta) +
\gamma u^{12} sin^6 (2 \theta) cos(6 [\phi_1 - \phi_2])
\end{equation}
\narrowtext
The phases obtained using eq.(\ref{Landau1}) then are:

\noindent
(I) $\beta_1>0$, $\beta_2>0$:
$\theta = \pi k/2$, corresponding to the symmetry class
$D'_6(E)$. This phase has a magnetic moment
along the direction of the $C_6$ axis
( the wave function $| 5/2 \rangle$ or $| - 5/2 \rangle$ ).

\noindent
(IIa) $\beta_1>0$, $-\beta_1<\beta_2<0$, $\gamma < 0$:
$\theta = \pi/4 + \pi k/2$, $\phi_1 - \phi_2 = \pi n/3$

\noindent
(IIb) $\beta_1>0$, $-\beta_1<\beta_2<0$, $\gamma > 0$:
$\theta = \pi/4 + \pi k/2$, $\phi_1 - \phi_2 = \pi/6 + \pi n/3$

The symmetry class of the wave function is in both cases $D'_2(E)$,
and the class for magnetic averages is $D_2(C_2)$.
A non-zero average moment in
the plane perpendicular to the six-fold axis
is expected for the magnetic class $D_2(C_2)$. However, with the wave
function $| - 5/2 \rangle + | 5/2 \rangle$ a symmetric 5-point correlator
is instead the lowest-order spin average:
\begin{equation}
T^{xxxxx} = - T^{xxxyy} = T^{xyyyy},
\end{equation}
with $T^{ijklm} = \langle J^i J^j J^k J^l J^m \rangle$.

In case of the representation $\tilde{E}_3$
all the phases are determined by the
fourth order terms of the Landau functional,
\widetext
\begin{equation}
F = - \alpha \tau u^2 + \beta_1 u^4 + \beta_2 u^4 sin^2 (2 \theta) +
\gamma u^4 sin^2 (2 \theta) cos( 2 [\phi_1-\phi_2]).
\end{equation}
\narrowtext

\noindent
(I) $\beta_1>0$, $\beta_2>|\gamma|$:
$\theta = \pi k/2$ ( the class $D'_6(\tilde{C}_3)_1$).
This phase  corresponds to the magnetic class $D_6(C_6)$
with the magnetic moment parallel to the
direction of the $C_6$ axis
(the wave functions $| 3/2 \rangle$ or $| - 3/2 \rangle$).

\noindent
(II) $\beta_1>0$, $|\gamma|-\beta_1<\beta_2<|\gamma|$, $\gamma > 0$:
$\theta = \pi/4 + \pi k/2$, $\phi_1 - \phi_2 = \pi/2 + \pi n$
(the class $D'_6(\tilde{C}_3)_2$).
The corresponding magnetic class is $D_6(D_3)$. The average magnetic
moment is prohibited by the symmetry of the magnetic class,
and the wave function $| 3/2 \rangle$ + $|-3/2 \rangle$ gives
the triple spin correlator of the form:
\begin{equation}
T^{xyy}= \langle J^x J^y J^y \rangle =
- T^{xxx} = - \langle J^x J^x J^x \rangle
\label{triple}
\end{equation}
as the magnetic order parameter.

\noindent
(III)  $\beta_1>0$, $|\gamma|-\beta_1<\beta_2<|\gamma|$, $\gamma < 0$:
$\theta = \pi/4 + \pi k/2$, $\phi_1 - \phi_2 = \pi n$
(the symmetry class is $D'_6(\tilde{C}_3)_3$).
The nonzero average for this wave function is
the triple spin correlator rotated
by an angle $\pi/2$ compared to the one in case (II).

We now apply the above analysis to the experimental situation
in UPt$_3$. Since no crystal field excitations have been seen
in  UPt$_3$\cite{UPt3}, mixing of
different configurations may be quite strong.
As a result, the small magnetic moment lying in the plane perpendicular
to the 6-fold axis in UPt$_3$ can be explained if the excited Kramers doublet
is $\tilde{E}_2$, with $J^z= \pm 5/2$, of the $5f^1$ configuration
(the symmetry class $D_2'(E)_2$ in Table \ref{table}).
Recall that the magnetic moment lying in the plane is not prohibited
by symmetry, but merely does
not appear on this particular ionic
Kramers doublet.
This moment cancellation is accidental, and small magnetic moment may
appear also as a result of admixture of the higher energy-level
multiplets. The
entropy loss is small due to the weakness of multipolar
interactions.

  The origin of the small magnetic moment in URu$_2$Si$_2$  must be
quite different.
As shown in Table \ref{table}, for the group $D_4'$ the
magnetic moment along the 4-fold axis is always present for the
MV order parameter Eq.(\ref{orpar}). In addition,
contrary to UPt$_3$, we found no accidental cancellation of the moment
due to a particular choice of the wave function for both $5f^1$
and $5f^3$ configuration.
Therefore, we assume that the transition takes place from
the mixed valency paramagnetic phase into a state where the coherent
function with $5f^2$ component overwhelms in Eq.(\ref{orpar}).
The crystal-field
excitations become well defined below the N\'eel transition
\cite{Mason:new}.
We propose that the real
matrix element between two singlet levels in the $5f^2$ configuration
prevails in the order parameter, so that the phase transition is
mostly structural in character.  This matrix element, according to
the symmetry analysis of
the Bragg reflexes for neutron scattering in the ordered state
\cite{Walker}, must have the symmetry
A$_z$ and double the unit cell.
Thus, the ground and excited integer spin states with the symmetries
$xy$ and $x^2-y^2$ or $A_1$ and $A_z$ both could be involved into the phase
transition. Such an assumption would account for
the large specific heat jump observed at the N\'eel point, and the intensive
crystal field excitations below $T_N$.
In order to explain the small magnetic moment,
we recall
that the valence of the uranium ion has a finite
(although, in this case, small)
probability to fluctuate, so that the coherent
wave function of the ionic subsystem below the phase transition has
a small admixture of different half-integer spin configuration
($5f^1$ or $5f^3$). Therefore,
the number of particles on the uranium sites is not
conserved, and the time-reversal symmetry below $T_N$ is broken. Then the
non-magnetic order component may be coupled to the
magnetic moment on the half-integer spin component:
$Q_z (|\eta_1|^2 - |\eta_2|^2)$.
Here $\eta_1 \psi_{\uparrow} + \eta_2 \psi_{\downarrow}$ is
the spinor admixture to the coherent wave function from the
$5f^1$ (or $5f^3$) configuration,
$Q_z=\langle \Psi|xy \rangle \langle x^2-y^2 |  \Psi \rangle$
(or $A_1 \otimes A_z$), is
the real matrix element for the non-magnetic component.
In other words, the smallness of the moment is then
ascribed to the valence fluctuations, reduced at the transition.

In Ref. \cite{Santini} the phase transition in URu$_2$Si$_2$
was assigned to the quadrupolar order parameter
($xy$ or $x^2-y^2$). Such a model, as noted above,
contradicts the neutron scattering Bragg peak pattern
\cite{Mason,Mason:new}, as shown in
\cite{Walker}.

To complete the above discussion, we make a few comments
concerning the above concept of phase transitions in the MV
intermetallides. As it was first suggested in Ref. \cite{Hirst},
valency fluctuations may be considered as excitations of an electron
above the Fermi energy changing the ionic configuration locally
(e. g. $5f^36d \ \leftrightarrow \ 5f^26d^2$ for the U-ion).
For the single impurity the generalized isotropic Anderson
model has been solved analytically \cite{Schlottmann}.  It was
shown that for such a model the resulting ground state could have
a small magnetic moment depending on the value of
parameters chosen. On the other hand, the two ionic configurations
are expected to differ in energy, at least, on the scale of the
spin-orbit coupling ( i. e. at least, a few tenths of eV). It
has been indicated by many authors (see Ref. \cite{Varma} and
references therein) that the further reduction of that scale
can be supplemented by the coupling to the lattice due
to the essential difference in ionic
sizes of two electronic configurations.

This coupling gives rise to large polaronic effects which, in particular,
strongly renormalize the matrix elements mixing both
configurations\cite{Sherrington}. We adopted above
this concept of the strong involvement of the lattice
variables into the phenomena related to the mixed valence compounds.
For the $5f$ orbitals having large spatial extent it could be useless
to describe the ionic state in
terms of single electron shells. The polaronic effects
may be responsible for electronic localization,
if the core Coulomb attraction at a given site is strong,
but cannot lead to the localized ground state by itself.
{}From this point of view the conduction electrons
scatter on the resonant complexes having both the energy
levels and their widths tuned by the interactions with the
quantized lattice displacements and the electroneutrality
condition. Such a regime may survive down to
lowest temperatures leading to the Fermi liquid Heavy Fermion
properties, in accordance with the scaling behavior of the single impurity
Anderson model. On the other hand, the lattice interactions might fix
the ultimate symmetry for these polaronic complexes without
destroying completely their coupling to the conduction sea.
We assume that this is
what happens in URu$_2$Si$_2$ at $T_N$. If
the above interpretation of the experimental data is correct,
and the structural component prevails, the smallness of the
average magnetic moments in URu$_2$Si$_2$ reflects the fact that the
width of the level (given by the mixing matrix element squared) is
small compared to the bare spin-orbital energy.

Like in the N\'eel state, where the average spin helps to describe (below
$T_N$)
the symmetry changes in otherwise strongly perturbed spin correlations,
the matrix element Eq.(\ref{orpar}) has served as implemental of this more
general
idea. The parameter Eq.(\ref{orpar}) describes the classification of the
electronic
degrees of freedom in the new coherent state of the lattice resulting from
trapping of electrons at T$_N$ by distortions around the uranium
ions. At the transition
the hybridization leaves the number of electrons not fixed at the local site,
i.e. the coherent electronic component may become a quantum mixture of the two
options.
Inside the trapped state it is the uranium potential which electrons see at
very short distances, but the truly local state  would not be possible without
the
local lattice adjustment.
If this view is correct, the lattice distortions are not separable
from the localization of electrons, though the electrons don't preserve their
number on each site. In this sense there is no room for questioning about the
statistics
of the mechanism which drives the transition utilizing the lattice degrees
of freedom and forming an electron state coherently for each trap.

   In summary, we relate the phenomenon of small magnetic moment
in some uranium compounds to the mixed valency features. We suggest that the
small magnetic moments in  UPt$_3$ and URu$_2$Si$_2$ are different in origin.
In UPt$_3$ the coherent wave
function corresponds to the
$5f^1$ configuration, while the magnetic moment vanishes for the
particular choice of the wave function. In case of URu$_2$Si$_2$ the
phase transition is primarily structural, with the small magnetic moment
resulting from a small admixture of a state of another valency.
One of us (L.P.G.) is grateful to W.J.L. Buyers, P. Coleman, Z.Fisk
and P. Fulde for
stimulating discussions. This work was supported (V.B.)
by the NSF Grant No. DMR91-20000 through
the science and Technology Center for Superconductivity
and (L.P.G.) by the NHMFL
through NSF cooperative agreement No. DMR-9016241
and the State of Florida.

\mediumtext
\begin{table}
\caption{Symmetry classes for the tetragonal and hexagonal groups.
The prime index stands for the double group.
Only the classes for representations
which are odd with respect to the $2 \pi$-rotation, $\tilde{E}$, are listed.
We use the usual notation $C_{\alpha n}^+(C_{\alpha n}^-)$ for the
clockwise (counterclockwise) rotation by the angle
$2 \pi/n$ with respect to the axis $\alpha$, as viewed from its positive
direction.
The elements of the group $\tilde{C}_3$ are: $\tilde{C}_3=\{E,
\tilde{C}_3^+, \tilde{C}_3^-\}$,
where $\tilde{C}_3^{\pm}=C_3^{\pm} \cdot \tilde{E}$.
We use rectangular coordinates for both $D_4'$ and $D_6'$ groups.
The $z$-direction corresponds to the four-fold or six-fold axis.
The $x$-direction is chosen to be perpendicular to the plane of the
Brillouin zone boundary.
The axes $a$ and $b$ for the group $D_4'$ are
along ($1$, $1$, $0$ ) and ( $1$, $-1$, $0$ ) directions, respectively.
The last index in the symmetry class column enumerates different phases.}
\begin{tabular}{ccccc}
Group & Representation & Symmetry & Generators & Magnetic    \\
      &                & Class    & of the Class & Order \\
\tableline
\tableline
$D_4'$ & $\tilde{E}_1$ & I. $D_4'(E)_1$ & $C_{4}^- e^{i \pi/4}$,
$C_{2y} e^{i \pi} R$ & $M_z$ \\
 & & IIa. $D_2'(E)_1$ & $C_{2z} R$, $C_{2b} e^{i \pi/2} R$ & $M_a$ \\
 & & IIb. $D_2'(E)_2$ & $C_{2z} R$, $C_{2y} e^{i \pi/2} R$ & $M_x$ \\
\tableline
 & $\tilde{E}_2$ & I. $D_4'(E)_2$ & $C_{4}^- e^{i 3 \pi/4}$,
$C_{2y} R$ & $M_z$ \\
 & & IIa. $D_2'(E)_1$ & $C_{2z} e^{i \pi} R$, $C_{2b} e^{-i \pi/2}$ & $M_b$ \\
 & & IIb. $D_2'(E)_2$ & $C_{2z} e^{i \pi} R$, $C_{2y} e^{- i \pi/2} R$ & $M_x$
\\
\tableline
\tableline
$D_6'$ & $\tilde{E}_1$ & I. $D_6'(E)_1$ & $C_6^- e^{i \pi/6}$,
$C_{2y} e^{i \pi/2} R$ & $M_z$ \\
 & & IIa. $D_2'(E)_1$ & $C_{2z} e^{i \pi/2} R$, $C_{2y} e^{- i \pi/2}$ & $M_y$
\\
 & & IIb. $D_2'(E)_2$ & $C_{2z} R$, $C_{2y} e^{i \pi/2} R$ & $M_x$ \\
\tableline
 & $\tilde{E}_2$ & I. $D_6'(E)_2$ & $C_{6}^- e^{i 5 \pi/6}$,
$C_{2y} e^{i \pi/2} R$ & $M_z$ \\
 & & IIa. $D_2'(E)_1$ & $C_{2z} e^{i \pi/2} R$, $C_{2y} e^{- i \pi/2}$ & $M_y$
\\
 & & IIb. $D_2'(E)_2$ & $C_{2z} R$, $C_{2y} e^{i \pi/2} R$ & $M_x$ \\
\tableline
& $\tilde{E}_3$ & I. $D_6'(\tilde{C}_3)_1$ & $C_{6}^- e^{i \pi/2}$,
$C_{2y} e^{i \pi/2} R$ & $M_z$ \\
 & & II. $D_6'(\tilde{C}_3)_2$ & $C_6^- R$, $C_{2y} e^{i \pi/2} R$ & $T_{xyy} =
- T_{xxx}$ \\
 & & III. $D_6'(\tilde{C}_3)_3$ & $C_6^- e^{i \pi/2} R$, $C_{2y} e^{- i \pi/2}$
 & $T_{yxx} = - T_{yyy}$ \\
\end{tabular}
\label{table}
\end{table}
\end{document}